\documentclass[11pt]{article}
\pdfoutput=1

\usepackage{a4wide}
\usepackage{Packages}

\usepackage{Definitions}

\usepackage{setspace}

\hypersetup{pdfauthor={Robbert Dijkgraaf, Domenico Orlando, Susanne
    Reffert},pdftitle={Relating Field Theories via Stochastic Quantization}}

\author{
  \begin{minipage}{.97\linewidth}
    \vspace{1cm}
    \begin{center}
      \begin{small}
        \textbf{Robbert Dijkgraaf}${}^{1,2}$, \textbf{Domenico Orlando}${}^3$ and \textbf{Susanne Reffert}${}^{3}$
      \end{small}
    \end{center}
    \vspace{1cm}
    \hspace{2cm}\begin{minipage}{.7\linewidth}
     {\it \begin{footnotesize}
     \begin{itemize}
	\item[${}^1$] KdV Institute for Mathematics, University of Amsterdam,\\
          Plantage Muidergracht 24, 1018 TV Amsterdam, The Netherlands.
	\item[${}^2$] Institute for Theoretical Physics, University of Amsterdam,\\
          Valckenierstraat 65, 1018 XE Amsterdam, The Netherlands. 
          \item[${}^3$] Institute for the Mathematics and Physics of the Universe, \\The University of Tokyo,
		Kashiwa-no-Ha 5-1-5, \\ Kashiwa-shi, 277-8568 Chiba, Japan.       \end{itemize}
     \end{footnotesize}}
    \end{minipage}
    \vspace{1cm}
  \end{minipage}
}

\date{}

\title{\vspace{1.5cm}
  \begin{huge}
    \textbf{Relating Field Theories via Stochastic Quantization}
  \end{huge}
}

\begin{document}

\numberwithin{equation}{section}

\begin{titlepage}
  \maketitle
  \thispagestyle{empty}

  \vspace{-14cm}
  \begin{flushright}
    ITFA-2009-08\\
    IPMU-09-0030
  \end{flushright}

  \vspace{14cm}

  \begin{center}
    \textsc{Abstract}\\
  \end{center}

  This note aims to subsume several apparently unrelated models under
  a common framework. Several examples of well--known quantum field
  theories are listed which are connected via stochastic
  quantization. We highlight the fact that the quantization method
  used to obtain the quantum crystal is a discrete analog of
  stochastic quantization. This model is of interest for string
  theory, since the (classical) melting crystal corner is related to
  the topological A--model. We outline several ideas for interpreting
  the quantum crystal on the string theory side of the correspondence,
  exploring interpretations in the Wheeler--De Witt framework and in
  terms of a non--Lorentz invariant limit of topological
  M--theory. \end{titlepage}
 
\onehalfspace

\tableofcontents



\section{Introduction}

In this note, we aim to connect several apparently unrelated theories
by showing that they are related through a common quantization scheme,
namely \emph{stochastic quantization}~\cite{Parisi:1980ys}. It finds
applications in field theory, the world of statistical and condensed
matter physics, and via correspondences also in string theory.

Models that can be regarded as resulting from stochastic quantization
all have a common structure. Starting from a (classical) model in $d$
dimensions one obtains a quantum model in $d+1$ dimensions as a Markov
process that converges at thermal equilibrium to the minima of the classical
action.  The partition function of the classical precursor model is
expressed as the norm square of a wave function, which is the ground
state of the theory in $d+1$ dimensions,
\begin{equation}
  Z^d_{\text{cl}}=\braket{\psi^{d+1}_{\text{ground}}|\psi^{d+1}_{\text{ground}}} \, .
\end{equation}

After summarizing the preliminaries of stochastic quantization, we
collect several examples of theories which are related by stochastic
quantization. The most basic and prototypical example is the one of
\emph{zero dimensional field theory}, whose stochastic quantization
leads straight to supersymmetric quantum mechanics.  In the case of
our next example, the \emph{free boson}, the naturally supersymmetric
description of the stochastically quantized theory provides a true
advantage over its prior treatment in the
literature~\cite{Ardonne:2003p1613}.  We show furthermore that the
quantization scheme employed to obtain the quantum crystal
in~\cite{Dijkgraaf:2008}, which is the same as the one used for the
quantum dimer model~\cite{Rokhsar1988}, is nothing else but a discrete
analog of stochastic quantization. In fact, both schemes are based on
an underlying \emph{Markov process}, which is a stochastic process for
which the probability for the system to be in a state at time $t$
depends only on the state at $t-1$. 

Even though stochastic quantization cannot be applied directly to the
\emph{gauged \textsc{wzw} model}, its partition function can be expressed as
the norm square of a 3--dimensional wave function, which is the ground
state of a strong coupling limit of topologically massive gauge theory
(Chern--Simons plus Yang Mills). This example thus belongs to the same
class of models.  The example of the \emph{quantum crystal} is
interesting from the point of view of string theory, since a
correspondence between the classical crystal melting configurations
and the topological string A--model has been
found~\cite{Okounkov:2003sp, Iqbal:2003ds}. The quantum crystal,
giving rise to a seven dimensional theory, suggests a connection to
topological M--theory. We outline some ideas concerning the
interpretation of the quantum crystal on the string theory side of the
correspondence, such as a \emph{Wheeler--De~Witt} interpretation of
the eigenvalue equation for the ground state.  Inspired by the
crystal/string correspondence, also the stochastic quantization of
\emph{K\"ahler gravity}~\cite{Bershadsky:1994sr} is sketched. Since
K\"ahler gravity is the target space description of the topological
A--model, its quantization should be related to the quantum crystal,
and via Hitchin's functional also to topological
M--theory~\cite{Dijkgraaf:2004te}. Given that quantum theories such as
those arising from the stochastic quantization of K\"ahler geometry
are all in need of regularization, the discrete framework of the
quantum crystal ultimately seems, if anything, most suited for
practical purposes.

\bigskip

The plan of this note is as follows. In Section~\ref{sec:introstoch},
we summarize the technique of stochastic quantization in the Langevin
and Fokker--Planck formulation and present its manifestly
supersymmetric form, as first given in~\cite{PARISI:1982p1560}. In
Section~\ref{sec:disc}, the discrete analog of the stochastic
quantization scheme is discussed. In Section~\ref{sec:Examples},
several examples of theories that are connected via stochastic
quantization are given. The prototypical example is the one of zero
dimensional field theory, discussed in
Section~\ref{sec:from-zero-dimensions}. The next example is the free
boson, see Section~\ref{sec:stoch-quant-boson}. In Sec.~\ref{sec:WZW}
we discuss the relation of the gauged \textsc{wzw} model to a strong coupling
limit of topologically massive gauge theory.  In Sec.~\ref{sec:qdm},
we briefly mention the quantum dimer model which is a precursor of the
quantum crystal. The latter is treated in Sec.~\ref{sec:qcrystal},
where also possible interpretations of the quantum system in string
theory are outlined. The last example concerns the stochastic
quantization of K\"ahler gravity, see Sec.~\ref{sec:Kgrav}.  A brief
summary and conclusion is given in Sec.~\ref{sec:conc}.


\section{Stochastic quantization revisited}
\label{sec:introstoch}

\emph{Stochastic quantization} is a quantization method for Euclidean field theories introduced in 1981 by Parisi and Wu~\cite{Parisi:1980ys}. It makes use of the fact that Euclidean Green's functions can be interpreted as correlation functions of a statistical system in equilibrium. A good overview over this quantization scheme including also subsequent work can be found in~\cite{Damgaard:1987p1516}. 

Starting from a $d$ dimensional Euclidean field theory, the field is
coupled to a white Gaussian noise which forces it to a random movement
on its manifold, a continuum analog of a random walk. This stochastic
system evolves along a new, fictitious, Euclidean time $t$. In the
limit $t\to\infty$, where thermal equilibrium is reached, the
$d$--dimensional correlation functions of the quantum field theory are
recovered, thus the original $d$--dimensional field theory has been
quantized. The equivalence of stochastic quantization and conventional
path integral quantization has been shown. When considered purely as a
way to quantize a $d$--dimensional field theory, the extra time
dimension is but a computational device which allows us to compute the
correlators of the $d$--dimensional quantum field theory after the
$\left( d + 1 \right)$--dimensional system has settled into
equilibrium. In this paper, we will study the $\left( d + 1
\right)$--dimensional quantum theory in its own right.

In the following, we will give a very brief introduction to the subject, referring the reader to the literature for the formal definitions of all the concepts used. We follow the notation used by~\cite{Damgaard:1987p1516}. 

\subsection{Langevin formulation}

A \emph{stochastic variable} $X$, defined by a range of values $x$ and a probability distribution $P(x)$ over these values, is generally used to describe the random fluctuations of a heat reservoir background. A \emph{stochastic process} is a process depending on the time $t$ and a stochastic variable. A \emph{Markov process} is a stochastic process in which the conditional probability for the system to be in a certain state at time $t$ only depends on its state at $t - \delta t$ and not on its earlier history.
The simplest example of a Markov process process is Brownian motion. Take a particle of mass $m$ moving in a liquid with friction coefficient $\alpha$. Its motion obeys the \emph{Langevin equation}
\begin{equation}
  \label{eq:langevin}
  m\frac{\mathrm d}{{\mathrm d}t} \vec{v}=-\alpha \vec{v}(t)+\vec{\eta}(t),
\end{equation}
where $\vec{\eta}(t)$ is the stochastic force vector representing the collisions between the particle and the molecules of the fluid. $\vec{\eta}(t)$ is a stochastic variable and can be assumed to have a Gaussian distribution.

Stochastic quantization of a Euclidean field theory works as follows. We supplement the field $\phi(x)$ with an extra time dimension $t$ (which must not be confused with the Euclidean time $x_0$). Then we demand that the time evolution of $\phi(x,t)$ obeys a stochastic differential equation such as the Langevin equation (\ref{eq:langevin}), which allows the relaxation to equilibrium:
\begin{equation}
  \label{eq:langevin_st}
  \frac{\partial\,\phi(x,t)}{\partial t}=-\frac{1}{2}\frac{\delta  S_{\text{cl}}}{\delta \phi}+\eta(x,t),
\end{equation}
with $S_{\text{cl}}$ the Euclidean action.
The correlations of $\eta$, which is a white Gaussian noise, are given by
\begin{align}\label{eq:corr_eta}
  \braket{\eta (x,t)} = 0 \, , && \braket{ \eta (x_1,t_1)
    \eta(x_2,t_t)} = 2 \delta ( t_1 - t_2)
  \delta^d ( x_1 - x_2 ) \, .
\end{align}
Equation (\ref{eq:langevin_st}) has to be solved given an initial
condition at $t=t_0$ leading to an $\eta$--dependent solution
$\phi_\eta(x,t)$. As a consequence, also $\phi_\eta(x,t)$ is now a stochastic variable. Its correlation functions are defined by
\begin{equation}
  \braket{\phi_\eta(x_1,t_1)\ldots\phi_\eta(x_k,t_k)}_\eta =
  \frac{\int\mathcal{D} \eta \, \exp\left(-\tfrac{1}{2} \int \di^d x
      \di t \, \eta^2(x,t) \right) \phi_\eta(x_1,t_1) \ldots
    \phi_\eta(x_k,t_k)} {\int\mathcal{D}
    \eta \, \exp\left(-\tfrac{1}{2}\int \di^d x\,\di t\,\eta^2(x,t)\right)}.
\end{equation}
The central points for stochastic quantization to make sense is  that
equilibrium is reached for $t\to \infty$, and that
\begin{equation}
  \lim_{t\to\infty} \braket{\phi_\eta(x_1,t) \ldots \phi_\eta(x_k,t)}_\eta = \braket{\phi(x_1) \ldots \phi(x_k)},
\end{equation}
\ie that the equal time correlators for $\phi_\eta$ tend to the corresponding quantum Green's functions.

\subsection{Fokker--Planck formulation}

The conditional probability distribution of the speed of a particle executing a Brownian motion depends on the initial conditions $P(v, t_0)$ and satisfies the \emph{Fokker--Planck} (FP)
equation:
\begin{equation}
  \label{eq:FP}
  \frac{\partial}{\partial t} P(v,t) = \frac{\partial}{\partial v}
  \left(v+\frac{\partial}{\partial v}\right)P(v,t) \, ,
\end{equation}
which can be derived from the Langevin equation (\ref{eq:langevin}) in
a standard way (after setting $ m = \alpha=1$). For stochastic quantization, the FP formulation can be used as an alternative approach.

Starting from (\ref{eq:langevin_st}), the evolution of the probability
$ P[\phi(x),t]$ in the extra time direction $ t$ is given by the FP
equation
\begin{equation} 
  \frac{\partial}{\partial t} P[\phi,t] = \int \di^d x \frac{\delta}{\delta     \phi(x,t)} \left( \frac{\delta S_{\text{cl}}}{\delta \phi(x,t)} + \frac{\delta}{\delta \phi(x,t)} \right)  P[ \phi,t] ,
\end{equation}
together with an initial condition.
One easily verifies that for $ t \to \infty$,
\begin{equation}
  \label{eq:equiP}
  \lim_{t \to \infty} P[ \phi, t] = P^{eq}[\phi] =       \frac{e^{-S_{\text{cl}}[\phi]}}{\int {\mathcal D} \phi \, e^{-S_{\text{cl}}[\phi]} }.
\end{equation}
The FP equation is a so--called \emph{master equation} (which means in
particular that its kernel is not Hermitian). To bring it into
Hamiltonian form we introduce
\begin{equation} 
  \psi[\phi, t] = P[\phi, t] e^{S_{\text{cl}}[\phi]/2}.
\end{equation}
This function $\psi$ satisfies the equation
\begin{equation} 
  \frac{\partial}{\partial t} \psi [\phi,t] = - \int\di^dx\, \HH[\phi] \, \psi[\phi,t] \, ,
\end{equation}
where 
\begin{equation}
  \label{eq:FPHamil}
  \HH[\phi] = - \frac{1}{2} \frac{\delta^2}{\delta \phi^2} + U[\phi],\ \ \text{and}\ \ U[\phi] = \frac{1}{8} \left( \frac{\delta S_{\text{cl}}}{\delta \phi} \right)^2 - \frac{1}{4}   \frac{\delta^2 S_{\text{cl}}}{\delta \phi^2} \, .
\end{equation}
The zero energy ground state for this Hamiltonian is 
\begin{equation}
  \label{eq:ground_st}
  \psi_0 [ \phi ] = e^{-S_{\text{cl}}[\phi]/2} \, .
\end{equation}
The corresponding Lagrangian can be obtained as usual by a Legendre transform:
\begin{equation}
  \label{eq:naive-lagrangian}
  \mathcal{L} [\phi, \dot \phi] = \frac{1}{2} {\dot\phi}^2 + U[\phi] \, .
\end{equation}

An alternative and more general way to find the quantum action consists in considering the partition function
\begin{equation}
  \mathcal{Z} = \braket{1}_\eta =  \int \mathcal{D} \eta \, e^{-\frac{1}{2} \int \eta^2 (x,t) \, dx dt}.
\end{equation}
Changing variables from $ \eta $ to $ \phi$, this becomes
\begin{equation}
  \label{eq:ZJacob}
  {\mathcal Z} = \int \mathcal{D} \phi \, \det \left( \frac{\delta \eta}{\delta \phi} \right) e^{-\frac{1}{2} \int \left( \frac{\partial}{\partial t} \phi(x,t) + \frac{1}{2} \frac{\delta S_{\text{cl}}[\phi]}{\delta \phi} \right)^2 dx dt}.
\end{equation}
The evaluation of the Jacobian requires some care. In particular,
one has to choose a direction for the time evolution, make use of the Stratanovich stochastic calculus (see~\cite{Arnold:1974}), and drop an
infinite constant. All these issues are neatly resolved in the supersymmetric
formulation presented in next section. By choosing propagation in the \emph{positive} time direction, one formally recovers
\begin{equation}
  \label{eq:Jacobian}
  \det \left( \frac{\delta \eta}{ \delta \phi} \right)= e^{\frac{1}{4} \int \frac{\delta^2 S_{\text{cl}}}{\delta \phi^2} dx dt} \, .
\end{equation}
Plugging the Jacobian back into~(\ref{eq:ZJacob}) and neglecting a
total derivative term $\dot \phi \delta S = \dot S$, one finds
\begin{equation}
  \mathcal{Z} = \int D \phi \, e^{- \int \mathcal{L} [\phi, \dot \phi] \, dx dt},
\end{equation}
where
\begin{equation}
  \label{eq:bosonic-stochastic-action}
  \mathcal{L} [\phi, \dot \phi] = \frac{1}{2} {\dot \phi}^2+ \frac{1}{8} \left( \frac{\delta S_{\text{cl}}}{\delta \phi} \right)^2 - \frac{1}{4} \frac{\delta^2 S_{\text{cl}}}{\delta \phi^2} \, ,
\end{equation}
as in Eq.~(\ref{eq:naive-lagrangian}).

\subsection{Supersymmetric formulation}

It was shown in~\cite{PARISI:1982p1560} that any theory obtained by
stochastic quantization admits supersymmetry with respect to the
fictitious time direction $t$. In fact, already the classical Brownian
motion in one dimension corresponds to supersymmetric quantum
mechanics. This supersymmetry is linked to the existence of the
so--called \emph{Nicolai map}. One of the useful consequences of this
supersymmetry which is already encoded in the Langevin equation, are
Ward identities for the Green's functions.

The fermionic superpartners of $\phi$ can be introduced as a
calculational device to expand the Jacobian determinant
(\ref{eq:Jacobian}) as a fermionic path integral:
\begin{equation}
  \det \left( \frac{\delta \eta}{\delta \phi} \right) = \det \left(
    \frac{\partial}{\partial t} + \frac{1}{2} \frac{\delta^2
      S_{\text{cl}}}{\delta \phi^2}\right) = \int \mathcal{D} \psi \mathcal{D} \bar \psi \, e^{\int dx dt\, \bar \psi \left( \frac{\partial}{\partial t} + \frac{1}{2} \frac{\delta^2 S_{\text{cl}}}{\delta \phi^2} \right)\psi}.
\end{equation}
The FP Lagrangian becomes
\begin{equation}
  \mathcal{L}_{FP} [\phi, \dot \phi, \psi, \bar \psi] = \frac{1}{2} {\dot \phi}^2 + \frac{1}{8} \left( \frac{\delta S_{\text{cl}}}{\delta \phi} \right)^2 - \bar \psi \left( \frac{\partial}{\partial t} + \frac{1}{2} \frac{\delta^2 S_{\text{cl}}}{\delta \phi^2} \right) \psi.
\end{equation}
In the scalar case, this can be rephrased in terms of superspace language. We introduce the superfield
\begin{equation} \Phi = \phi + \bar \theta \psi + \bar \psi \theta + \bar \theta \theta F,
\end{equation}
where $F$ is an auxiliary field needed to achieve closure of the supersymmetry algebra.
The super--covariant derivatives are given by
\begin{equation}\label{eq:supercovariant}
 D = \frac{\partial}{\partial \bar \theta} + \theta \frac{\partial}{\partial t},\quad\quad \overline D = \frac{\partial}{\partial \theta} + \bar \theta \frac{\partial}{\partial t} .
\end{equation}
With this, the action becomes
\begin{equation}\mathcal{L}_{FP} [ \Phi ] = S_{\text{cl}}[\Phi] + \Phi \overline D D \Phi,
\end{equation}
where $S_{\text{cl}}[\Phi]$ is the integral over the classical Lagrangian in terms of the superfield $\Phi$. Note in particular that
$S_{\text{cl}}[\Phi]$ now takes the role of a superpotential which is
a function of the superfield $\Phi$.  A word should be spent on the
interpretation of the supersymmetry in this type of problems. When
passing from the Langevin to the FP description, there are two
possible choices for the direction of the time propagation. They
correspond to two Hamiltonians, $ \mathcal{H}^- $ and $ \mathcal{H}^+
$. $ \mathcal{H}^-$ is the one we used; $ \mathcal{H}^+ $ has by
construction only strictly positive eigenvalues and hence does not
contribute to the $ t \to \infty $ dynamics where the
classical system is recovered. The partition function can be expressed as the
difference $Z = Z^+ - Z^-$ using the corresponding Lagrangians. For
some functions $ F[\phi]$ of the fields, the expectation values on the
two partition functions coincide and we find identities of the type
\begin{equation} \langle F \rangle_Z = \langle F \rangle_{Z^+} - \langle F \rangle_{Z^-} = 0,
\end{equation}
which can be interpreted as Ward identities in the supersymmetric formalism.

\bigskip

Consider now the special case of the action for a free field $ \phi$ which can be written in the form
\begin{equation}
  S_{\text{cl}}[ \phi ] = \braket{ \phi | L\, \phi} \, ,
\end{equation}
where $ \phi \in \mathscr{H}$, with $ \mathscr{H}$ being a Hilbert space with scalar product $\braket{\cdot | \cdot}$ and $ L$ a self--adjoint operator. The Langevin equation for $ \phi_t$ reads
\begin{equation}
  \dot \phi_t = - L\, \phi_t + \eta_t ,
\end{equation}
where $ \eta_t \in \mathscr{H}$ is a white Gaussian noise. The corresponding partition function is
\begin{equation} 
  \mathcal{Z} = \int \mathcal{D} \eta_t \, \exp \left[ \frac{1}{2} \int d t \, \braket{\eta_t| \eta_t} \right].
\end{equation}
Changing the integration variable,
\begin{equation}
  {\mathcal Z} = \int \mathcal{D} \phi_t \, \det \left( \frac{\delta \eta_t }{\delta \phi_t } \right) \exp \left[ \frac{1}{2} \int d t \, \left\| \eta_t  \right\|^2 \right].
\end{equation}
We introduce the Grassmann fields $ \psi, \bar \psi \in \mathscr{H}$ to express the determinant in a fermionic representation,
\begin{equation}
  \det \left( \frac{\delta \eta_t}{\delta \phi_t} \right) = \int \mathcal{D} \bar \psi_t       \mathcal{D} \psi_t \, \exp \left[ - \int dt \, \braket{\bar \psi_t | \frac{\delta \eta_t}{\delta           \phi_t} \psi_t } \right] \, .
\end{equation}
Now the final action takes the form
\begin{equation} 
  S_{\text{q}}[\phi_t,\dot\phi_t,\psi_t,\dot\psi_t ]= \frac{1}{2} \left\| \dot \phi + L \phi \right\|^2 - \braket{ \bar \psi | \dot \psi + L \psi } \, .
\end{equation}
This same action can be written in a manifestly supersymmetric form as
\begin{equation}
  S_{\text{q}}[\Phi] = \int d \bar \theta d \theta \, S_{\text{susy}} = \frac{1}{2} \int d \bar \theta d \theta \, \braket{ \overline D \Phi | D \Phi} + \braket{\Phi | L\, \Phi} \, .
\end{equation}

\subsection{Discrete analog}\label{sec:disc}

It is possible to quantize discrete stochastic models in a manner analogous to
stochastic quantization as discussed above.  The dynamics of any discrete
stochastic model is described by the \emph{master equation}
\begin{equation}
  \label{eq:master}
  \frac{\mathrm{d}}{\mathrm{d}t}\, P_\alpha(t) = \sum_{\substack{\beta\\\beta\neq\alpha}}\left(W_{\alpha\beta} P_\beta(t) - W_{\beta \alpha} P_\alpha(t)\right),
\end{equation}
where $P_\alpha(t)$ is the probability to be in configuration
$\alpha$ at time $t$, and $W_{\alpha \beta}$ is the
\emph{transition rate} to state $\alpha$ if the system is in state
$\beta$. 
We choose the following transition rates:
\begin{equation}
  W_{\alpha \beta} = C_{\alpha \beta}\, e^{- g \left( \overline{\mathcal H}( \alpha ) - \overline{\mathcal H} (\beta)  \right)/ 2}  \, ,
\end{equation}
where $g$ is a coupling constant, $\overline{\mathcal H}(\alpha)$ is
the classical Hamiltonian evaluated on the configuration $\alpha$
and 
$C_{\alpha \beta}$ is the adjacency matrix of the state
graph. One can verify that the (unique) stationary distribution for
$P_\alpha $ is
\begin{equation}\label{eq:P0}
  P_\alpha^{(0)} = \frac{1}{Z}\, e^{-g \overline{\mathcal H}(\alpha)} \, ,  
\end{equation}
where 
\begin{equation}
  Z = \sum_\alpha e^{-g \overline{\mathcal H}(\alpha)} \, .  
\end{equation}
This is equivalent to saying that the system is in a Boltzmann
distribution with classical energy $\overline{\mathcal H}$.  
One can easily verify not only that $P_\alpha^{(0)}$ describes an
equilibrium state, but that it also satisfies the \emph{detailed balance}
condition
\begin{equation}
  \label{eq:detailedBalance}
  W_{\beta\alpha}P^{(0)}_\alpha=W_{\alpha \beta}P^{(0)}_\beta.
\end{equation}
Detailed balance implies the Markov property for a stochastic process.

It is customary to define the exit rate from state $\alpha $ as 
\begin{equation}
  W_{\alpha \alpha} = - \sum_{\beta \neq \alpha }
  W_{\beta \alpha} = - \sum_{\beta \neq \alpha }   C_{\alpha \beta}
  e^{-g \left( \overline{\mathcal H}( \beta ) - \overline{\mathcal H}
      (\alpha )  \right)/ 2} \, .
\end{equation}
In this way, the evolution can be described by the vector equation 
\begin{equation}
  \frac{\di }{\di t } \mathbf{P} (t) = \mathbf{W} \mathbf{P} (t) \, ,
\end{equation}
where $\mathbf{W}$ is the matrix with entries $W_{\alpha \beta}$. The
stationarity condition becomes
\begin{equation}
  \mathbf{W} \mathbf{P}^{(0)} = 0 \, .
\end{equation}
Instead of using the matrix $\mathbf{W}$, one can define the
symmetrized version with entries
\begin{equation}
  \widetilde{W}_{\alpha \beta} = \left( P_\alpha^{(0)} \right)^{-1/2} W_{\alpha \beta}  \left( P_\beta^{(0)} \right)^{-1/2} 
\end{equation}
(no summation implied), which is now interpreted as the Hamiltonian of
the dynamical system.  Explicitly, one finds
\begin{equation}
  \label{eq:Discrete-Laplacian}
  \begin{cases}
    \widetilde{W}_{\alpha \beta} = C_{\alpha \beta} & \text{if $\alpha \neq \beta$} \\
    \widetilde{W}_{\alpha \alpha } = W_{\alpha \alpha } \, .
  \end{cases}
\end{equation}
Note that this expression coincides with the definition for the Laplacian on a directed graph given in \cite{Chung:2005}.

Given the characteristic equation $\widetilde{W}_{\alpha \beta} \tilde
\phi_\beta^{(\lambda)} = \lambda \tilde \phi_{\alpha }^{(\lambda)} $
one can verify that the ground state is now represented by the vector with components
\begin{equation}
  \label{eq:ground_disc}
  \tilde\phi_\alpha^{(0)} = e^{-g \overline{\mathcal H}(\alpha)/2} \, .
\end{equation}

This whole construction can now be interpreted as a quantization
procedure by introducing a Hilbert space $\mathscr{H}$ generated by
vectors labelled by the states $\alpha $. Now $\widetilde{W}$ is
interpreted as a Hamiltonian operator, and the evolution equation for
$\tilde \phi \in \mathscr{H}$ as a real time Schr\"odinger equation.

\bigskip

The analogy to the stochastic quantization for a continuous theory
becomes obvious once one considers the following points.
\begin{itemize}
\item The Markov property lies at the root of both quantization
  schemes.
\item Just as Eq. (\ref{eq:master}), also the Fokker--Planck equation
  (\ref{eq:FP}) is a master equation.
\item The stationary distribution (\ref{eq:P0}) has the same form as
  the equilibrium probability in Eq.~(\ref{eq:equiP}).
\item The procedure of symmetrizing the matrix $\mathbf W$ is
  analogous to the one of bringing the FP equation into a
  Schr\"odinger--like form leading to the Hamiltonian
  (\ref{eq:FPHamil}).
\end{itemize}
Last but not least,
\begin{itemize}
\item the ground states (\ref{eq:ground_st}) and (\ref{eq:ground_disc}) are the same.
\end{itemize}


\section{Examples}
\label{sec:Examples}

This section forms the main body of this article and gives several
examples of models which are related by the framework of stochastic quantization. We
cite some illustrative examples from the literature along with
examples which we present here for the first time.

We start by considering the prototypical example of a zero dimensional
field theory which becomes supersymmetric quantum mechanics after
quantization. We next discuss the simple but rich example of the stochastic
quantization of a bosonic field. Our treatment offers a
reinterpretation of the discussion of the quantum Lifshitz model
in~\cite{Ardonne:2003p1613} and allows the generalization beyond the
free case.
We next discuss the relation of the gauged \textsc{wzw} model to the strong
coupling limit of topologically massive gauge theory, which belongs to
the same class of models even though stochastic quantization cannot be
applied directly.  The resulting limit of topologically massive gauge
theory is, unlike the pure Chern--Simons theory, dynamical.

Then we move on to the discrete examples of the quantum dimer model
and the quantum crystal. Here we stress the point that this discrete
quantization scheme is the analog of stochastic quantization and explore
possible interpretations of the results in terms of  string theory.
Finally, we sketch the quantization of the six--dimensional K\"ahler
gravity action.

Recently, theories with anisotropic scaling which can be thought of as stemming from a stochastic quantization process have appeared in the literature in the context of gauge theory, membrane actions, and non---Lorentz invariant gravity~\cite{Horava:2008jf, Horava:2008ih, Horava:2009uw}.

Another example which we are
not discussing here deserves being mentioned, namely the connection
between $d=3$ Chern--Simons theory and $d=4$ topological Yang---Mills
theory via stochastic quantization, as given in~\cite{Baulieu:1988jw}.

\subsection{From zero dimensions to supersymmetric quantum mechanics}
\label{sec:from-zero-dimensions}

The simplest example of stochastic quantization is obtained by
considering zero dimensional quantum field theory. In this case,
the field is a map from a point $P$ to the real line, $x:P \to
\setR$, \emph{i.e.} a variable. The action is a function of this
variable, $S_{\text{cl}}= S_{\text{cl}}(x)$, and the classical partition
function is given by the integral
\begin{equation}
  Z_{\text{cl}} = \int_\setR \di x \; e^{-S_{\text{cl}} (x)} \, .  
\end{equation}
The stochastic quantization is performed by adding a time direction
and promoting $x$ to a function of time, $x: \setR \to \setR,\ \ t
\mapsto x(t)$. We can now impose the Langevin equation given in Eq.~(\ref{eq:langevin_st}):
\begin{equation}
  \frac{\di}{\di t}x(t) = - \frac{1}{2} \frac{\di}{\di x} S_{\text{cl}} [x(t)] + \eta (t) \, ,
\end{equation}
where $\eta (t)$ is a white Gaussian noise.  This system corresponds to a one dimensional Brownian motion. Using the procedure
outlined above, we can write the quantum partition function as:
\begin{equation}
  \mathcal{Z} = \int \mathcal{D} x(t) \; e^{-\frac{1}{2} \int \di t \, \eta(t)^2 } = \int \mathcal{D} x(t) \det \left[\frac{\partial \eta(t)}{\partial x(t)}\right] e^{-\frac{1}{2} \int \di t \, \left( \dot x^2 + \frac{1}{2} S^\prime_{\text{cl}} [x(t)]\right)^2} \, ,  
\end{equation}
whence we can directly read the quantum action, which is given precisely
by the usual supersymmetric quantum mechanics,
\begin{equation}
  S_{\text{q}} [ x, \psi, \bar \psi ] = \int \di t \, \left[ \frac{1}{2} \dot x(t)^2 + \frac{1}{8} \left(S_{\text{cl}}^\prime[x(t)]  \right)^2 - \bar \psi(t) \dot \psi(t) - \frac{1}{2} S_{\text{cl}}^{\prime \prime} [x(t)] \bar \psi(t) \psi (t) \right] \, ,  
\end{equation}
where $\psi $ and $\bar \psi$ describe a complex fermion. The
corresponding Hamiltonian density is given by
\begin{equation}
  \HH [ \pi, x, \psi, \bar \psi ] = \frac{1}{2} \pi^2 + \frac{1}{8} \left( S^\prime_{\text{cl}} [x]\right)^2 + \frac{1}{2} S_{\text{cl}}^{\prime \prime} [x] \comm{\bar \psi, \psi} \, ,
\end{equation}
where $\pi$ is the conjugate momentum to $x(t)$. Being a
supersymmetric theory, it makes sense to introduce the charges
\begin{gather}
  Q = \bar \psi ( \frac{\partial}{\partial x} + \frac{1}{2} S_{\text{cl}}^\prime [x] ) \, ,\\
  \overline Q =  \psi ( -  \frac{\partial}{\partial x} + \frac{1}{2} S_{\text{cl}}^\prime [x] ) \, ,
\end{gather}
such that the Hamiltonian is obtained as the anticommutator
\begin{equation}
  \HH = \frac{1}{2} \acomm{Q, \overline Q} \, .
\end{equation}
The ground state of the theory is annihilated by both charges and can be written as:
\begin{equation}
  \psi_0[x] = e^{-S_{\text{cl}}[x]/2} \, .
\end{equation}
As expected, the classical partition function factorizes,
\begin{equation}
  Z_{\text{cl}} = \int_\setR \di x \; e^{-S_{\text{cl}} (x)} = \int \di x \, \abs{ \psi_0 [x] }^2 = \braket{\psi_0|\psi_0}_{L^2}\, .  
\end{equation}

\subsection{Stochastic quantization of a bosonic field}\label{sec:boson}
\label{sec:stoch-quant-boson}

As a slightly more complicated example, we treat next the stochastic quantization of a free boson.
Consider the action for a free boson in $d$ Euclidean dimensions,
\begin{equation}
  S_{\text{cl}}^d[\varphi] = \frac{\kappa}{2} \int \di^d x \, \left[  \partial_i
  \varphi (x^i)\, \partial^i \varphi(x^i) \right] \hspace{2cm} i=1,2, \dots, d \, .
\end{equation}
The Langevin equation, describing the evolution in the fictitious time
$t$ is given by
\begin{equation}
  \partial_t \varphi ( t,  x^i) = \frac{\kappa}{2} \, \partial_i \partial^i \varphi
  (t, x^i) + \eta ( t, x^i)  \,.
\end{equation}
where $\eta (t, x^i)$ is a white Gaussian noise, \emph{i.e.} a
stochastic variable whose second momentum is the only one that is non--vanishing, see (\ref{eq:corr_eta}).
In this way, $\varphi (t, x^i)$ itself becomes a stochastic variable and the
expectation value of any functional $F[\varphi]$ is obtained by
averaging over the noise:
\begin{equation}
  \braket{F[\phi]}_\eta = \frac{1}{\mathcal{Z}} \int \mathcal{D} \eta \,
  F[\varphi] e^{-\frac{1}{2} \int \di t\,\di^{d} x \, \eta
    (t, x^i)^2} \, ,
\end{equation}
where the partition function is defined by
\begin{equation}
  \mathcal{Z} = \int \mathcal{D} \eta \, e^{-\frac{1}{2} \int \di t\,\di^{d} x \, \eta
    (t,x^i)^2} \, .
\end{equation}
It is convenient to change the integration variable from $\eta $ to
$\varphi$.  The expression becomes:
\begin{equation}
  \mathcal{Z} = \int \mathcal{D} \varphi \, \left. \det \left[ \frac{\delta
      \eta}{\delta \varphi} \right] \right|_{ \eta = \partial_t \varphi ( t,x^i) + \kappa \, \partial_i \partial^i \varphi} e^{-\frac{1}{2} \int \di  t\,\di^{d} x \,
  \left( \partial_t \varphi - \frac{\kappa}{2} \, \partial_i \partial^i \varphi \right)^2} \, .
\end{equation}
The Jacobian is easily expressed by introducing two fermionic fields
$\psi (t,x^i)$ and $\bar \psi(t,x^i)$ such that
\begin{equation}
  \left. \det \left[ \frac{\delta
      \eta}{\delta \varphi} \right] \right|_{ \eta = \partial_t
  \varphi ( t,x^i) + \kappa \, \partial_i \partial^i \varphi}  = \int \mathcal{D} \psi \mathcal{D} \bar \psi
   \, e^{- \int \di t\,\di^{d} x \, \bar \psi(t,x^i) \left( \partial_t -
       \frac{\kappa}{2} \, \partial_i \partial^i \right) \psi(t,x^i)} \, .
\end{equation}
In this way we can directly read off the $\left( d + 1
\right)$--dimensional action:
\begin{equation}
\label{eq:quantized-free-boson}
  S^{d+1}_{\text{q}} [ \varphi, \psi, \bar \psi]  = - \int \di t\,\di^{d}x \,
  \left[ \frac{1}{2} \left( \partial_t \varphi \right)^2 +
    \frac{\kappa^2}{8} \left( \partial_i \partial^i \varphi \right)^2
    - \bar \psi \left( \partial_t -
      \frac{\kappa}{2} \, \partial_i \partial^i \right) \psi \right] \, .
\end{equation}
We can rewrite this result by introducing the superfield
$\Phi$ defined by
\begin{equation}
  \Phi (t,x^i, \theta, \bar \theta) = \varphi (t,x^i) + \bar \theta
  \psi (t,x^i ) + \bar \psi(t,x^i) \theta + \bar \theta \theta
  F(t,x^i) \, ,  
\end{equation}
where $\theta $ and $\bar \theta$ are Grassmann variables and
$F(t,x^i)$ is an auxiliary bosonic field. One finds that
\begin{equation}
  S^{d+1}_q [ \varphi, \psi, \bar \psi]= \int \di \bar \theta \di \theta\, \di t\,\di^{d} x \, \left[  D \Phi \overline D
    \Phi + \mathcal{L}_{\text{cl}} [ \Phi ] \right] = \int \di \bar \theta
  \di \theta\,\di t\, \di^{d} x \, \left[ D \Phi \overline D
    \Phi + \frac{\kappa}{2} \partial_i \Phi \partial^i \,\Phi \right] \, ,  
\end{equation}
where $D$ and $\overline D$ are the super--covariant derivatives (\ref{eq:supercovariant}).
Note that here, as remarked before, $\mathcal{L}_{\text{cl}}$ takes the role of a superpotential.
In this form the action is manifestly covariant under the
supersymmetric variations
\begin{align}
  \delta_{\epsilon} \varphi = \bar \epsilon \psi - \bar \psi \epsilon, && \delta_{\epsilon} F = \bar \epsilon \dot \psi + \dot {\bar \psi} \epsilon ,\\
  \delta_{\epsilon} \psi = \epsilon \left( \dot \varphi + F \right), && \delta_{\epsilon} \bar \psi = \bar \epsilon \left( \dot \varphi - F \right). 
\end{align}

There are different possible approaches to obtain the Hamiltonian
description. One possibility consists in starting with the
Fokker--Planck equation. Alternatively, one can use the supersymmetric
structure and define the two supercharges
\begin{align}
  Q = \bar \psi \left( \imath\Pi_\varphi - \frac{\kappa}{2}
    \, \partial_i \partial^i \varphi \right), && \overline Q = \psi \left( - \imath
    \Pi_\varphi - \frac{\kappa}{2}
    \, \partial_i \partial^i \varphi \right) \, ,
\end{align}
where $\Pi_\varphi = - \imath \frac{\delta}{\delta \varphi}$ is the conjugate momentum to $\varphi$. One can
easily see that $\set{Q, Q} =\set {\overline Q, \overline Q} = 0$ and
\begin{equation}
  \set{ Q, \overline Q} = 2 \HH \, ,  
\end{equation}
where $\HH$ is the Hamiltonian operator obtained by a Legendre
transformation of the action $S_{\text{q}}^{d+1}$ in
Eq.~(\ref{eq:quantized-free-boson}). Using the expressions for $Q$
and $\overline Q$, one can verify that the (bosonic) ground state, 
\begin{equation}
  \ket{\Psi_0} = e^{ -\frac{\kappa}{4} \int \di^d x \, \partial_i
    \varphi \partial^i \varphi} \ket{0} \, ,
\end{equation}
where $\ket{0}$ is the vacuum annihilated by $\psi$, is in fact a
supersymmetric ground state, since
\begin{equation}
  Q \ket{\Psi_0} = \overline Q \ket{\Psi_0} = 0 \, .  
\end{equation}
The bosonic part of the action in Eq.~\eqref{eq:quantized-free-boson}
has already been conjectured in~\cite{Henley2003} and studied in~\cite{Ardonne:2003p1613}.  
The case $d=2$ is particularly interesting
since the ground state expectation values of the $d=3$ quantum theory
can be evaluated as correlators of a conformal field theory.

The supersymmetric formulation presented here has the advantage that due to the presence of the fermionic part, both $Q$ and $\overline Q$ annihilate the bosonic ground state, which was not the case for the quantum Lifshitz model discussed in~\cite{Ardonne:2003p1613}. In their case,  the contribution to the Hamiltonian coming from $Q\overline Q$ was interpreted as a UV divergent zero--point energy and had to be subtracted.
Using the stochastic quantization formalism thus allows us to deal with more general problems where the
boson evolves in a potential and is described by an action of the type
\begin{equation}
  S_{\text{cl}}^d[\varphi] = \int \di^d x \, \left[ \frac{\kappa}{2} \partial_i
  \varphi \partial^i \varphi + V [\varphi] \right] \, .
\end{equation}
Take for example a massive theory,
\begin{equation}
  V[\varphi] = \frac{m^2}{2} \varphi^2 \,  .
\end{equation}
The $\left(d +1 \right)$--dimensional action is found to be
\begin{multline}
  S_{\text{q}}^{d+1} [ \varphi, \psi, \bar \psi]= \int \di^{d+1} x \, \left[ \frac{1}{2}
  \left(\partial_t \varphi \right)^2 + \frac{\kappa^2}{8}
  \left( \partial_i \partial^i \varphi \right)^2 + \frac{\kappa m^2}{4} \partial_i
  \varphi \partial^i \varphi + \frac{m^4}{8} \varphi^2 -\right.\\
  \left.- \bar \psi
  \left(\partial_t + \frac{m^2}{2} - \frac{\kappa}{2} \, \partial_i \partial^i \right) \psi \right] \, .
\end{multline}
This action provides a supersymmetric generalization of the Landau
free energy expression used to describe the Lifshitz points in the
study of liquid crystals~\cite{Chaikin:1995} (a Lifshitz point occurs
where a high--temperature disordered phase, a spatially uniform ordered
phase and a spatially modulated phase meet).
The supersymmetric charges are found to be
\begin{align}
  Q = \bar \psi \left( \frac{\delta}{\delta \varphi} - \frac{\kappa}{2}
    \, \partial_i \partial^i \varphi + \frac{m^2}{2} \varphi \right) \, , && \overline Q = \psi
  \left( - \frac{\delta}{\delta \varphi} - \frac{\kappa}{2}
    \, \partial_i \partial^i \varphi + \frac{m^2}{2} \varphi \right) \, .
\end{align}
Another interesting example is provided by starting with a sine--Gordon type action.
\begin{equation}
  S_{\text{cl}}^d [\varphi] = \int \di^d x \,  \left[\frac{\kappa}{2} \partial_i
  \varphi \partial^i \varphi + \frac{\lambda}{2 \pi} \cos ( 2 \pi
  \varphi)  \right] \, .  
\end{equation}
In this case, the quantization yields
\begin{multline}
  S_{\text{q}}^{d+1} [ \varphi, \psi, \bar \psi] = \int \di^{d+1}x \, \left[\frac{1}{2} \left( \partial_t \varphi \right)^2
  + \frac{1}{2} \left( \kappa \,\partial_i \partial^i \varphi + \lambda
    \sin (2 \pi \varphi) \right)^2 -\right.\\
    \left.- \bar \psi \left( \partial_t - \frac{\kappa}{2}
    \kappa \, \partial_i \partial^i - \pi \lambda \cos (2 \pi
    \varphi) \right) \psi  \right] \, ,
\end{multline}
and the $Q$ and $\overline Q$ operators read:
\begin{align}
  Q = \bar \psi \left( \frac{\delta}{\delta \varphi} - \frac{\kappa}{2}
    \, \partial_i \partial^i \varphi - \frac{\lambda}{2} \sin (2 \pi \varphi) \right) \, , && \overline Q = \psi
  \left( - \frac{\delta}{\delta \varphi} - \frac{\kappa}{2}
    \, \partial_i \partial^i \varphi - \frac{\lambda}{2} \sin (2 \pi \varphi)  \right) \, .
\end{align}
Since for a precise choice of $\lambda$ and $\kappa$, the sine--Gordon
action corresponds to a massive Dirac fermion, we expect this action to
be related to the continuum limit of the quantum crystal. 
It was argued in~\cite{Ardonne:2003p1613} that this action
should give rise to a mass gap which is in agreement with
the results in~\cite{Dijkgraaf:2008} concerning the discrete model. It
is likely that the supersymmetric structure we identified here will help in gaining a better understanding of the properties of the system away from the Lifshitz point represented by the free
boson (quantum dimer).

\subsection{From the gauged \textsc{wzw} model to topologically massive gauge theory}
\label{sec:WZW}

As the next easiest model to study, let us now consider the gauged \textsc{wzw}
model. This example differs a little from the others since we are not
using stochastic quantization directly. But it belongs to this
collection because the two theories (in $d$, resp. $d+1$ dimensions)
are related by the classical partition function of one being the norm
square of the ground state wave function of the other. This is the
common property of \emph{all} the examples discussed here.

We first summarize the \textsc{wzw} model very briefly and refer the reader to the literature for details. The basic field $g$ of the \textsc{wzw} model is a map from a 2d Riemann surface $\Sigma$ to a compact Lie group $G$. Its basic functional is the action 
\begin{equation}
  I[g] = -\frac{1}{8\pi}\int_\Sigma \mathrm{d}^2\sigma\sqrt\rho \rho^{ij}\Tr(g^{-1}\partial_ig\cdot g^{-1}\partial_jg)-i\,\Gamma[g],
\end{equation}
where $\rho$ is a metric on $\Sigma$ and $\Gamma$ is the Wess--Zumino term,
\begin{equation}
  \Gamma[g]=\frac{1}{12\pi}\int_B\mathrm d^3\sigma\epsilon^{ijk}\Tr g^{-1}\partial_i\,g\cdot g^{-1}\partial_j\, g\cdot g^{-1}\partial_k \,g,
\end{equation}
where $B$ is a 3--manifold such that $\partial B=\Sigma$. The main
property of the \textsc{wzw} model that we will make use of in the following is
the existence of a \emph{holomorphic factorization}. This means that
the partition function $Z$ can be expressed as a square,
$Z=\braket{f|f}$, of a holomorphic section $f$ of a flat vector bundle
over moduli space.

As it was already noted in~\cite{Ardonne:2003p1613}, a direct
stochastic quantization of the \textsc{wzw} action is not possible because of
the presence of the \textsc{wz} term\footnote{The \textsc{wz} term, being
  imaginary, would not contribute to the modulus
  square of the action. As a result, one would find at thermal equilibrium,
  $t \to \infty$, that the system converges to the principal chiral model
  instead of the expected \textsc{wzw} model.}. Instead we follow Witten's
treatment~\cite{Witten:1991mm}, where the \textsc{wzw} model is gauged in order
to derive the existence of the holomorphic factorization property. In
the gauged model, $g$ becomes a section of a bundle $X\to\Sigma$ with
fiber $G$ and structure group $G_L\times G_R$. We take $A$ to be a
connection on such a bundle.  Let us now define
\begin{equation}
  I[g,A] = I[g] + \frac{1}{2\pi}\int_\Sigma \mathrm{d}^2 z \Tr A_{\bar z}g^{-1}\partial_zg-\frac{1}{4\pi}\int_\Sigma \mathrm{d}^2z\Tr A_{\bar z}A_z.
\end{equation}
Now we can formally define a functional of A, 
\begin{equation}\label{eq:wavefn}
  \Psi[A]=\int \mathcal{D}g\, e^{-k\,I[g,A]}.
\end{equation}
$\Psi[A]$ obeys two key equations,
\begin{align}
  \label{eq:propertyA}
  \left(\frac{\delta}{\delta A_z}-\frac{k}{4\pi}A_{\bar z}\right) \Psi[A] &= 0,\\
  \left(D_{\bar z}\frac{\delta}{\delta A_{\bar z}}+\frac{k}{4\pi}D_{\bar z}A_z-\frac{k}{2\pi}F_{\bar z z}\right)\Psi[A] &=0,
  \label{eq:propertyB}
\end{align}
where the covariant derivative is defined by
\begin{equation}
  D_i u=\partial_i u + \comm{ A_i, u }.
\end{equation}
These two conditions mean that $\Psi[A]$ is a \emph{holomorphic section} and
\emph{gauge invariant}. This can be interpreted to mean that $\Psi[A]$ is a
physical state, \ie a wave function, of $2+1$ dimensional
Chern--Simons theory, and thus relates the \textsc{wzw} model to Chern--Simons
theory. In \cite{Witten:1991mm} it is shown that the $\Psi[A]$ we
constructed in Eq.~(\ref{eq:wavefn}) is indeed the holomorphic section
that squares to the partition function,
\begin{equation}
  Z(\Sigma) = \frac{1}{\mathrm{vol}(\hat G)} \int \mathcal{D} A \, \overline{\Psi[A]} \Psi [A] = \norm{\Psi}^2.
\end{equation}
We recognize here the by now familiar structure of a theory resulting from stochastic
quantization: the partition function of a classical theory in $d$
dimensions (in this case \textsc{wzw} and $d=2$) is expressed as the square of
the ground state wave function of a quantum theory in $d+1$
dimensions. All we need to know now is the precise theory whose
Hamiltonian annihilates $\Psi[A]$. The corresponding theory was described
in~\cite{Grignani:1996ft} and is a strong coupling limit of
topologically massive gauge theory, whose action is given by
\begin{equation}
  S = - \int \di^3 x \, \Tr [ \frac{1}{2} F_{0i} F^{0i}] + \frac{k}{4
    \pi} \int \di x \, \Tr [ A dA - \frac{2}{3} \imath A^3 ] \, .
\end{equation}
We would now like to point out another similarity with stochastic
quantization. Interestingly enough, the Chern--Simons term is
reminiscent of the total derivative term $\frac{dS}{dt}$ that we
dropped in the derivation of the action in
Eq.~(\ref{eq:bosonic-stochastic-action}). When considering a
3--manifold $Y = \Sigma \times [0,T]$, on the one hand the
contribution of $\int_Y \di x \di t \frac{\di S}{\di t}$ is
$\int_{\Sigma = \partial Y} \left( S(T) - S(0) \right)$, and on the
other, it is a known fact that on such a manifold, the \textsc{cs}
action is equivalent to the (chiral) \textsc{wzw} model on the
boundary $\partial Y$~\cite{Moore:1989yh,Elitzur:1989nr}.\\
Note moreover that in the strong coupling limit we consider, the
magnetic component of the Yang--Mills term drops out. As a result, the
Lagrangian, as it is usually the case for stochastic quantization, is
not Poincar\'e invariant. Canonically quantizing this model on $\Sigma
\times \setR $ (where $\Sigma $ is the Riemann sphere) in the Weyl
gauge $A_0 = 0$, one obtains the Hamiltonian\footnote{If instead of
  taking the strong coupling limit we had considered the complete
  Yang--Mills term, this would have resulted in an extra term in the
  Hamiltonian,
  \begin{equation}
    \HH = \int \di^2 x  \, \Tr[E (x^i)^2 + B(x_i)^2 ] \, , 
  \end{equation}
  which does not lead to the ground state we expect.}
\begin{equation}
  \label{eq:MTGT-Hamiltonian}
  \HH = \int d^2 x \, \Tr [ E(x^i)^2 ] \, , \hspace{2em}  i = 1,2 \, , 
\end{equation}
where
\begin{equation}
  E^{a,i}(x) = \Pi^{a,i} (x) - \frac{k}{8 \pi} \epsilon^{ij}A^a_j(x) \, ,  
\end{equation}
and $\Pi^{a,i} (x^j)$ is the conjugate momentum of $A^{a}_i(x^j)$:
\begin{equation}
  \comm{A_i^a (x), \Pi_j^{b} (y)} = \imath \, \delta_{ij} \delta^{ab}
  \delta(x-y) \, . 
\end{equation}
In this formalism, we can now understand  the two conditions in
Eq.~(\ref{eq:propertyA}) and Eq.~(\ref{eq:propertyB}):
\begin{itemize}
\item Since $A_0$ appears in the action as a Lagrange
  multiplier, the gauge choice $A_0 = 0$ is implemented by imposing
  Eq.~(\ref{eq:propertyB}) as the constraint that the state $\ket{\Psi}$
  be physical.
\item Equation~(\ref{eq:propertyA}) is now understood as $\Psi [A]$
  being the ground state annihilated by the operator $E^a$:
  \begin{equation}
    E^a \ \Psi[A] = \left( \frac{2}{\imath} \frac{\delta}{\delta A_z^a} -
      \frac{\imath k}{8\pi} A_{\bar z}^a \right) \Psi[A] = 0 \, .
  \end{equation}
  (Here, we passed to complex coordinates where $A_z = A_1 + \imath
  A_2$ and $A_{\bar z} = A_1 - \imath A_2$).
\end{itemize}

Being a first order equation, we would like to argue that the
last relation can be understood as the annihilation of the ground
state in a supersymmetric theory. To do so, we introduce the following
operators:
\begin{gather}
  Q^a(x) = \bar \chi(x)  \left( \frac{2}{\imath} \frac{\delta}{\delta A_z^a} -
    \frac{\imath k}{8\pi} A_{\bar z}^a \right) = \bar \chi (x) \chi E(x) \, , \\
  \overline Q^a(x) = \left( - \frac{2}{\imath} \frac{\delta}{\delta A_{\bar z}^a} +
    \frac{\imath k}{8\pi} A_z^a \right) \chi(x) = \bar E(x) \chi(x) \, ,
\end{gather}
where $\chi $ is a complex fermion satisfying
\begin{equation}
  \acomm{ \chi (x), \bar \chi (y) } = \delta (x-y) \, .
\end{equation}
The anticommutator between $Q^a $ and $\overline Q^a$ yields
\begin{equation}
  \acomm{Q^a(x), \overline Q^a (y) } = \left( \bar E^a(x) E^a(x) +
    \frac{k }{2 \pi }\bar \chi(x) \chi (x) \right) \delta (x-y) \, ,
\end{equation}
where we used the commutation relation
\begin{equation}
  \comm{ E^a(x), \bar E^b (y) } = \frac{k}{2 \pi} \delta^{ab}
  \delta(x-y) \, .
\end{equation}
The anticommutator is a supersymmetric extension of the Hamiltonian
of topologically massive gauge theory,  see
Eq.~(\ref{eq:MTGT-Hamiltonian}), which only contains the bosonic
part. Using the charges $Q^a(x),\  \overline Q^a(x)$, one can directly derive the supersymmetry
transformations for the fields $A^a$ and $\chi$, which are given by
\begin{align}
  \delta_\epsilon A^a = \comm{\epsilon Q^b + \bar \epsilon \overline Q^b, A^a},\\
  \delta_\epsilon \chi = \comm{\epsilon Q^b + \bar \epsilon \overline Q^b, \chi}  \, .  
\end{align}
Going to the Lagrangian formalism, the system is described by the action
\begin{equation}
  S = - \int \di^3 x \, \Tr [ \frac{1}{2} F_{0i} F^{0i}] + \bar \chi \dot \chi + \frac{k}{2\pi} \bar \chi \chi + \frac{k}{4
    \pi} \int \di x \, \Tr [ A dA - \frac{2}{3} \imath A^3 ] \, .
\end{equation}
This can be seen as a strong coupling limit of the supersymmetric
topologically massive gauge theory studied in \cite{Deser:1981wh}.

\subsection{The quantum dimer model}
\label{sec:qdm}

This example comes from the world of solid state physics.
The quantum dimer model is a so--called \emph{resonating valence bond}
model and was first introduced by Rokhsar and
Kivelson~\cite{Rokhsar1988} as a candidate model for high temperature
superconductivity. It starts from the \emph{dimer model}\footnote{For the detailed definitions see for example~\cite{Kenyon2}.}, a lattice model
living on a bipartite two--dimensional graph. Dimers live on the edges
of the graph and each vertex can only be touched by one dimer. In the
dimer model, the number of so--called \emph{perfect matchings} is
counted. The basic move in the dimer model is a \emph{plaquette flip}, in
which the dimers of a fully occupied plaquette of the lattice are each
turned by one position.  The quantization of the dimer model is an
example of a discrete analog of stochastic quantization as discussed
in Section~\ref{sec:disc}.  Its Hilbert space is spanned by vectors in
one--to--one correspondence with the perfect matchings and its quantum
Hamiltonian can be expressed as
\begin{equation}
  \label{eq:QHamil_dimer}
  \HH =  - J \left(\sum\ketbra{\blacksquare}{\square} +  \ketbra{\square}{\blacksquare} +  \sum   \ketbra{ \square}{\square} + \ketbra{ \blacksquare}{\blacksquare}\right) \, ,
\end{equation}
where the black and white squares correspond to the two states in
which a fully occupied plaquette can be. The sum runs over all
flippable plaquettes of a given perfect matching. The first two terms are
kinetic and flip a plaquette either to the left or the right, the last
two terms are potential terms and count the number of plaquette
moves. The ground state corresponds to the equally weighted sum over
all perfect matchings. 

In the case of the dimer model on the hexagonal lattice, there exists a one--to--one map
between perfect matchings and the configurations of an idealized
three--dimensional crystal corner. The Hamiltonian in
Eq.~(\ref{eq:QHamil_dimer}) can be generalized to describe a (Markov) growth process as detailed in the
next section.

\subsection{Quantum crystal melting}\label{sec:qcrystal}

Let us now discuss the example of quantizing the melting crystal
corner~\cite{Dijkgraaf:2008}, which is interesting from the point of
view of string theory. This is as well an example of a discrete analog
of stochastic quantization as discussed in Section~\ref{sec:disc}.

\bigskip

The melting crystal corner is often mapped to the problem
of stacking cubes in an empty corner of 3D space. The growth rules are
the following. A cube can be added to a configuration if three of its sides will touch
either the wall or other cubes. This leads to a minimum energy
configuration without any free--standing cubes. The partition function
of the melting crystal corner takes the following form:
\begin{equation}
  \label{eq:melting}
  Z = \sum_{3d\ \text{partitions}}q^{\text{\# boxes}}=\prod_{n=1}^\infty\, \frac{1}{\left( 1 - q^n \right)^n} \, ,
\end{equation}
where the rightmost expression is the so--called \emph{MacMahon function}. 

Crystal melting is an extension of the dimer model discussed in the
last section in the following sense. Via a rhombus tiling, the
configurations of the dimer model on the hexagonal lattice are in
one--to--one correspondence with the configurations of the melting
crystal corner. While in the dimer model, only the overall number of
configurations are counted, the model of the melting crystal corner
contains more information, it also keeps track of the number of cubes
of each configuration. The quantum Hamiltonian of the melting crystal
can be expressed as
\begin{equation}\label{eq:QHamil}
  \HH =  - J \left(\sum\ketbra{\blacksquare}{\square} +  \ketbra{\square}{\blacksquare} +  \sum  \sqrt{q}  \ketbra{ \square}{\square} + \frac{1}{\sqrt{q}} \ketbra{ \blacksquare}{\blacksquare}\right) \, ,
\end{equation}
which acts on the Hilbert space generated by the orthonormal basis of the
configurations $\ket{\alpha}$. The sum runs over all places in a given configuration
where a cube can be added or removed. The ket $\ket{ \blacksquare}$
represents a place where a cube can be removed,
while $\ket{ \square}$ represents a place where a cube can be
added. Once one takes the limit $q\to 1$ in (\ref{eq:QHamil}), one
recovers the Hamiltonian of the quantum dimer model.  The ground state
is in fact the unique zero energy ground state, fulfilling
\begin{equation}
  \label{eq:gs}
  \HH \ket{\text{ground}}=0,
\end{equation} 
and has the form
\begin{equation}
  \label{eq:ground1}
  \ket{\text{ground}} = \sum_\alpha  q^{N(\alpha)/2} \ket{\alpha } \, .
\end{equation}
Note the normalization for the wave function,\begin{equation}
  \braket{\text{ground} | \text{ground}} = \sum_\alpha q^{\text{\# boxes}} = Z \, .  
\end{equation}
Once we identify 
\begin{equation}
  \label{eq:q}
  q = e^{- g_s},
\end{equation}
it turns out that the closed string partition function of the A--model
with the Calabi--Yau $X={\mathbb C}^3$ as target space corresponds
exactly to the partition function of the melting crystal
corner~(\ref{eq:melting})~\cite{Okounkov:2003sp, Iqbal:2003ds}. For
the crystal, the above identification~(\ref{eq:q}) translates into a
lattice spacing of $g_s$. The crystal melting configurations can be
roughly mapped to (multiple) blow--ups $\hat X$ of the original
Calabi--Yau $X$. Moreover, in~\cite{Ooguri:2008yb} it was shown that a
crystalline structure similar to the one corresponding to
$\mathbb{C}^3$ which we used, can be constructed for \emph{any} toric Calabi--Yau
manifold, based on the corresponding quiver gauge theory. The
partition function of this crystalline model will then correspond to
the partition function of the topological A--model on this toric
Calabi--Yau modulo possible wall crossings (see
\emph{e.g.}~\cite{Szendroi:2007nu}). This means that also our quantization
scheme can be generalized to any toric Calabi--Yau. All that is needed
is the state graph whose nodes  are crystal configurations. Its
arrows mark which configurations are related by adding or removing an
atom. The quantization procedure does not depend on the details
of the crystal.  Consider the state graph for the
melting of a crystal made of $A$ different atoms. To each
configuration $\alpha $ we associate a weight
\begin{equation}
  w (\alpha  ) = \prod_{i = 1}^A q_i^{n_i (\alpha )} \, , 
\end{equation}
where $0 < q_i < 1,\hspace{1em} i = 1, \dots, A $ are parameters and
$n_i (\alpha )$ is the number of atoms of species $i$ in the
configuration $\alpha $. Then the Laplacian $\widetilde{W}$ in
Eq.~(\ref{eq:Discrete-Laplacian}) becomes the matrix with entries
\begin{equation}
  \widetilde{W}_{\alpha \beta} =
  \begin{cases}
    1 & \text{$\alpha $ and $\beta$ are neighbours,} \\
    - \displaystyle{\sum_{\text{$\gamma$ neighbour of $\alpha $ }}} \sqrt{\frac{w(\gamma )}{w(\alpha  )}} & \text{if $\alpha  = \beta$,} \\
    0 & \text{otherwise.}
  \end{cases}
\end{equation}
One can verify that the vector
\begin{equation}
  \ket{\text{ground}} = \sum_\alpha  \left( \prod_{i=1}^A q_i^{n_i(\alpha )/2} \right) \ket{\alpha }     
\end{equation}
is a zero energy ground state for $\widetilde{W}$ and its norm square
reproduces the classical crystal melting partition function
\begin{equation}
  \braket{\text{ground} | \text{ground} } = \sum_\alpha  w(\alpha ) = Z \sim Z^{\text{top}}\quad\text{(modulo wall crossings)}\, .    
\end{equation}

Let us now consider possible interpretations of the quantized crystal in terms of string theory.
By adding a time evolution to the statistical system, we created a
quantum theory with one dimension more than the system we started from. 

From the string theory point of view, this means that instead of the
six dimensional topological A--model, we should be looking at a seven
dimensional theory.  An immediate candidate would be topological
M--theory, or an effective version thereof. Given that the new time
dimension is not treated on the same footing as the original six
dimensions, we cannot be looking at M--theory itself, but rather at a
non--Lorentz invariant limit. Unfortunately, topological M--theory has
not yet been clearly defined. Its status is best discussed
in~\cite{Dijkgraaf:2004te}, where its connection to Hitchin's
functionals is described.

Given that the classical configurations of the melting crystal are
related to geometries, it would be interesting to try to find a
completely geometric interpretation of the
Hamiltonian~(\ref{eq:QHamil}), which is the Laplacian on the graph of
states (geometries). Its kinetic term creates "neighboring"
geometries from a given configuration. It would be interesting to see
if also the potential term has a precise geometric meaning.

\bigskip

Concerning the interpretation of the quantum crystal, another
observation can be made that we consider to be one of the central
conceptual points of this paper. If we identify the three--dimensional
crystal configurations with multiple blow--ups of $\setC^3$, we find
that the Hilbert space generated by the configurations has a natural
interpretation as \emph{mini--superspace} and the ground state can be
seen as a \emph{wavefunction of the Universe}.

Looking at formula~(\ref{eq:gs}),
\begin{equation}
\label{eq:our-condition}
  \HH \ket{ \Psi_0} = 0  \,,
\end{equation}
one is strongly reminded of the \emph{Wheeler--De~Witt equation}. The
\textsc{wdw} equation also has the form $\HH_G \Psi=0$, where $\HH_G$ is the
Hamiltonian associated to the Einstein--Hilbert action, and $\Psi$ is
the wave function of the universe. The \textsc{wdw} equation is in fact nothing
else but a zero energy Schr\"odinger equation and the crucial
information needed is the initial quantum state which is then
evolved. The wave function $\Psi$ is defined on \emph{superspace}, an
infinite dimensional space of all possible geometries and matter
configurations. In practice, one usually works in the finite
dimensional \emph{mini--superspace}, in which all but a few degrees of
freedom are frozen out.

The sum over classical crystal melting configurations corresponds on
the A--model side to a sum over "quantum K\"ahler geometries", or to
be more precise, a sum over ideal sheaves, which are torsion--free
sheaves with vanishing first Chern class~\cite{Iqbal:2003ds}. In this
framework, the mini--superspace we are looking at is thus the moduli
space of all ideal sheaves ${\mathcal M}^{sheaf}$ over the original
manifold $X$.

In our case however, it is not obvious how to find the natural
classical gravity theory whose quantization leads to $\HH$ (\ie the analog of
the Einstein--Hilbert action the usual \textsc{wdw} equation descends
from). Let us therefore concentrate on the meaning of $\HH$. As we
have stressed already, $\HH$ is the Laplacian on the space of
configurations. The condition in
Eq.~(\ref{eq:our-condition}) can be understood as requiring $\Psi_0$
to be a harmonic function in mini--superspace. This is strongly
reminiscent of the results in~\cite{Moncrief:1989dx} concerning
quantum gravity in $2+1$ dimensions. In this case, the
Einstein--Hilbert gravity turns into a Hamiltonian system on
Teichm\"uller space after the ADM reduction. In particular, for genus
$g=1$, the \textsc{wdw} equation takes the form
\begin{equation}
  \label{eq:sqrt}
  \sqrt{\Delta} \ket{ \Psi } = 0  \,,
\end{equation}
where $\Delta$ is the Laplacian on the torus moduli space.
Two observations can be made about this:
\begin{itemize}
\item (quantum) gravity in $2+1$ dimensions is special because there
  are no propagating degrees of freedom. It is therefore not surprising that
  this case is very close to our construction which is related to
  topological strings.
\item The square root in~(\ref{eq:sqrt}) is reminiscent of the fact
  that in supersymmetric systems the ground state is found by imposing
  $Q \ \Psi_0 = 0$. The operator $Q$ is in some sense close to being
  the ``square root'' of the Hamiltonian, since $\HH=\tfrac{1}{2}\{Q,
  \overline Q\}$. Also the discrete quantum crystal can be
  supersymmetrized (by enlarging the space of states) and the ground
  state then satisfies a first order difference equation as opposed to
  the second order equation~(\ref{eq:our-condition}).  Apart from
  this, Eq.~(\ref{eq:sqrt}) results in the same ground state as
  Eq.~(\ref{eq:our-condition}).
\end{itemize}

\bigskip

There is yet another way of looking at the classical crystal
configurations, as adopted in~\cite{Ooguri:2008yb}, namely as BPS
bound states of $D$--branes ($D0$ and $D2$ branes bound to a single
$D6$ brane). Removing an atom of the crystal corresponds here to
adding a specific combination of $D0$ and $D2$ brane charges. In this
picture, the quantum Hamiltonian~(\ref{eq:QHamil}) jumps thus between configurations
with different D--brane charges.

\bigskip

In the above paragraphs, we have outlined possible directions to
pursue in order to arrive at a meaningful interpretation of the quantum crystal on the
string theory side of the correspondence. We leave a more thorough investigation of
these ideas for future research.

\subsection{K\"ahler gravity?}
\label{sec:Kgrav}

In the previous section we proposed a possible interpretation of our
quantum crystal construction in terms of a Wheeler--De~Witt
equation. To make this connection more precise we would need an
appropriate theory of gravity to quantize. A possible candidate is a
6--dimensional form theory of gravity, namely K\"ahler gravity, which describes variations of the K\"ahler structure on a
complex manifold $M$~\cite{Bershadsky:1994sr}. Also this example is
interesting from the point of view of string theory, since it provides
the target space description of the topological A--model. Following
the logic in~\cite{Okounkov:2003sp, Iqbal:2003ds}, its stochastic
quantization should provide a theory that corresponds to (or is at
least in the same universality class and thus shares important
properties with) the quantum crystal. A relation of this theory to
topological M--theory is therefore to be expected.  Here we sketch a
possible approach leading to the stochastic quantization of this
theory.

The action of Kähler gravity is given by
\begin{equation}
 S_{\text{Kaehler}}[K]= \int_M   \frac{1}{2} K \frac{1}{d^{c \dagger}} d K + \frac{1}{3} K \wedge K \wedge K \, ,
\end{equation}
where $K$ is a variation of the (complexified) Kähler form
on $M$, and $d^c = \partial - \bar{\partial}$.
The Kähler gravity action is
invariant under gauge transformations of the form
\begin{equation}
 \delta_{\alpha} K = \di \alpha - \di^{c \dagger} (K \wedge \alpha) \, ,
\end{equation}
where $\alpha$ is a 1-form on $M$, such that $d^{c \dagger} \alpha =0$.
The equations of motion for Kähler gravity theory have the form
\begin{equation}
  \di K + \di^{c \dagger} (K \wedge K) = 0.
\end{equation}
We can decompose $K$ into massless and massive modes,
\begin{align}
  K = x + \di^{c \dagger} \gamma, && x \in H^{1,1} (M,\setC),
\end{align}
where $x \in H^{1,1}(M,\setC)$ represents the Kähler moduli, which are
not integrated over, and $\gamma \in \Omega^3(M)$ contains the massive
modes of $K$.  Using the above decomposition we can write the Kähler gravity
action without non--local terms as follows:
\begin{equation}
  S_{\text {Kaehler}} [x,\gamma] = \int_M
  \frac{1}{2} d \gamma \wedge d^{c \dagger} \gamma
  + \frac{1}{3} K \wedge K \wedge K \, . 
\end{equation}
Lagrangian D--branes of the A--model are charged
under $\gamma$, implying that these branes are sources for $K$ and hence
modify the integral of $K$ on 2--cycles which link them.  This also
implies that the partition function of the A--model depends
non--perturbatively on the choice of a cohomology class in $H^3(M)$ as
well as on $x\in H^2(M)$. 

Let us now implement the stochastic quantization procedure in the
supersymmetric approach. The seven--dimensional theory can be written
as
\begin{equation}
  S_{\text{7 d}} = \int \di \bar \theta \di \theta \di t \int_M \overline D \Gamma \wedge \star D \Gamma + S_{\text{Kaehler}} [ x, \Gamma]  \, , 
\end{equation}
where $D$ and $\bar D$ are the super--covariant derivatives, $\star$ is
the Hodge star on the 6d manifold, and $\Gamma$ is the superfield
defined by
\begin{equation}
  \Gamma = \gamma + \bar \theta \psi + \bar \psi \theta + \bar \theta \theta F  ,
\end{equation}
where $\psi, \bar \psi, F$ are all three--forms in space.  Note that
$x$ contains the Kähler moduli and is not promoted to a
superfield. Just as we have seen above, the classical action now takes
the role of a superpotential.

\bigskip

The above derivation is purely formal: the properties of the seven
dimensional action have to be understood, and it is not obvious that
the stochastic quantization of Kähler gravity is the most direct way
towards a description of the seven dimensional topological
M--theory. On the other hand we believe that some of the general
features that we encountered in the previous examples (quantum theory
in $d+1$ dimensions whose ground state reproduces the partition
function of the classical theory in $d$ dimensions, supersymmetry in
the extra dimension) are to be expected for such a theory.



\section{Conclusions}\label{sec:conc}

In this article we have discussed a rather widespread, even though often
unrecognized, scheme to relate a \emph{classical} to a \emph{quantum} field
theory. In condensed matter physics, this method of
quantization gave rise to the (discrete) quantum
dimer model and its generalization, the quantum crystal. In the continuous case,
this approach goes under the name of \emph{stochastic quantization}. In
topological field theories it has appeared for example in relating the
\textsc{wzw} model to the Chern--Simons action.
 Its main features are the following.
\begin{enumerate}
\item The classical theory lives on a $d$--dimensional manifold
  $\Sigma$, while the $\left( d + 1 \right)$--dimension\-al (quantum)
  theory lives on the manifold $Y = \Sigma \times [0,T]$.
\item In the Hamiltonian description, the quantum theory admits a
  zero--energy ground state wavefunction $\Psi_0$. Its norm square
  reproduces the classical partition function $Z_{\text{cl}}$:
  \begin{equation}
    \braket{ \Psi_0 | \Psi_0 } = Z_{\text{cl}} \, .    
  \end{equation}
\item The quantum theory can be interpreted as a \emph{Markov process} that
  at equilibrium converges to the minima of the classical action. In
  this sense, the quantum theory describes
  the same physics in the $T \to \infty$ limit as the classical theory.
\item There is a natural supersymmetric structure in the extra
  dimension.
\end{enumerate}

Based on point number 2 above, one might even venture to say that
every theory that admits a \emph{holomorphic factorization} of its
partition function might have a $\left(d+1 \right)$--dimensional partner whose
ground state square results in the partition function of the original theory.

\bigskip
In this note, we have supplied
a number of examples of models which are as such widely known, while
their relation to each other by stochastic quantization on the other
hand is largely unrecognized. 

We have done two things:
\begin{enumerate}
\item Having recognized a common principle in seemingly unrelated
  models, we are able to overcome some of the problems
  encountered in previous works, notably by making use of the inherent
  supersymmetry.
\item Since this quantization scheme underlies the
  quantum crystal, which is deeply related to topological strings, we
  would like to propose it as an approach towards topological
  M--theory.
\end{enumerate}

We have sketched some ideas for interpreting the quantum crystal on
the string theory side of the correspondence, exploring
interpretations in the \textsc{wdw} framework and in terms of a non--Lorentz
invariant limit of topological M--theory.  Much work remains to be
done in both directions and a more thorough study of these ideas is
left for future research.


\subsection*{Acknowledgements}

It is a pleasure to thank Simeon Hellerman, Hirosi Ooguri, Cumrun Vafa, and Erik Verlinde for discussions. 
D.O. would like to thank the University of Amsterdam for hospitality. S.R. would like to thank the University of Neuch\^atel for hospitality.
Furthermore, the authors would like to thank the V. Simons Workshop in Stony Brook for hospitality, where this work was initiated.

The research of R.D. was supported by a NWO Spinoza grant and the FOM program "String Theory and Quantum Gravity."
The research of D.O. and S.R. was supported by the World Premier International Research Center Initiative (WPI Initiative), MEXT, Japan.



\bibliography{DimerReferences}

\end{document}